\documentclass[preprint,12pt]{elsarticle}

\usepackage{graphicx}
\usepackage{dcolumn}
\usepackage{bm}

\usepackage[utf8]{inputenc}
\usepackage[T1]{fontenc}
\usepackage{mathptmx}
\usepackage{amssymb}

\usepackage[colorlinks=false, linkcolor=black, citecolor=black, urlcolor=black]{hyperref}
\usepackage{natbib}

\newcommand{\onlinecite}{\cite}

\journal{Journal of Alloys and Compounds}

\begin{document}
\begin{frontmatter}

\title{Reliable thermodynamic estimators for \\ screening caloric materials} 

\author[address1]{Nikolai~A. Zarkevich \corref{correspondingauthor}}
\cortext[correspondingauthor]{Corresponding author} 
\ead{zarkev@ameslab.gov} 

\author[address1,address2]{Duane~D. Johnson}
\ead{ddj@iastate.edu, ddj@ameslab.gov}
\address[address1]{Ames Laboratory, U.S. Department of Energy,  Ames, Iowa 50011-3020, USA}
\address[address2]{Materials Science \& Engineering, Iowa State University, Ames, Iowa 50011-2300, USA}

\date{\today}
 
\begin{abstract}
Reversible, diffusionless, first-order solid-solid phase transitions accompanied by caloric effects are critical for applications in the solid-state cooling and heat-pumping devices.  
Accelerated discovery of caloric materials requires reliable but faster estimators for predictions and high-throughput screening of system-specific dominant caloric contributions.
We assess reliability of the computational methods that provide  thermodynamic properties in relevant solid phases at or near a phase transition.
We test the methods using the well-studied B2 FeRh alloy as a ``fruit fly'' in such a materials genome discovery, as it  exhibits a meta\-magnetic transition which generates multicaloric (magneto-, elasto-, and baro-caloric) responses.
For lattice entropy contributions, we find that the commonly-used linear-response and small-displacement phonon methods are invalid 
near instabilities that arise from the anharmonicity of atomic potentials, 
and we offer a more reliable and precise method for calculating lattice entropy at a fixed temperature. 
Then, we apply a set of reliable methods and estimators to the metamagnetic transition in FeRh (predicted $346 \pm 12 \,$K, observed $353 \pm 1 \,$K) 
and calculate the associated caloric properties, such as isothermal entropy and isentropic temperature changes.  
\end{abstract}

\begin{keyword}
caloric, thermodynamic, metamagnetic, phase transformation, FeRh. 
\end{keyword}

\end{frontmatter}


\section{\label{Introduction}Introduction}

Solid-state caloric devices have a potential to save vast amounts of electricity  \cite{PhilTransRoySocA374p20150305y2016,ADMA23n7p821y2011,IntJRefrigeration31n6p945y2008,Ioffe1954,Giauque1927,Debye1926}.
However, predicting thermodynamics in a caloric material can be challenging \cite{Cazin1875}, as near the phase transformation -- where caloric effects are induced -- the system is on the edge of stability, often with multiple instabilities competing.  
Hence, thermodynamic estimators need a serious assessment before applications to caloric systems \cite{JPhysD51n2p024002y2018},
 or for use in high-throughput screening supplemented using databases and machine-learning techniques. 

{\par} The caloric effect is typically quantified by the isothermal entropy change $\Delta S_T$ and associated isentropic temperature change $\Delta  T_S$ 
at the phase transition at a critical temperature $T_c$. 
But these are not the only important quantities.  
Others include the enthalpy change $\Delta H$ at a fixed pressure $P$ or temperature $T$ (importantly, $\Delta H_P \ne  \Delta H_T$), 
the hysteresis width, dependences of $T_c$ on composition and external fields, etc. 
Thus, a search for a good caloric material involves simultaneous optimization of multiple parameters. 
For their accurate prediction, it is important to take into account several contributing physical effects, using multi-physics, multi-parameter modeling.  
On the other hand, quick estimates of the lower and upper bounds allow fast rejection, needed for the high-throughput materials screening. 

{\par} Our key goal here is to test the reliability of various (often commonly used) methods and to validate our results with those that are measured. 
The overarching need is a set of reliable, and preferably fast, estimators for thermodynamic quantities for screening, especially for desired outliers -- say, materials with a large caloric response. 
 Such materials, however, have electronic (including magnetic) and structural instabilities,  in which case the vibrational contributions are often not harmonic; and yet quasi\-harmonic phonon methods are commonly used. 

{\par}  To analyze and test methods and estimates, we use the multicaloric FeRh system. 
With its chemical simplicity and well-studied meta\-magnetic transition, FeRh serves as a wonderful ``fruit fly'', or test system, in the materials genome discovery of better caloric systems \cite{JPhysD51n7p070201Manosa}. 
However, a long-studied material is not necessarily well understood; there is a continued controversy among the directly measured and indirectly assessed experimental data, as we discuss.

{\par}  
Interestingly, FeRh \cite{JChemPhys35p1904y1961,Abrahamson1966,JAP37n3p1257y1966,JAP38p1263y1967,PSSB20n1pK25y1967} and NiTi austenite \cite{TiNi1963,PRB90p060102y2014,PRL113p265701y2014,APL101n7p073904y2012} have the same nominal chemical B2 structure (CsCl, $Pm\bar{3}m$ space group, see Fig.~\ref{fig1}) and exhibit a large caloric effect \cite{Nikitin1990p363,Annaorazov1996,ActaMat106p15y2016,JETP36n1p105y1973,Pugacheva1994p731,PRB50p4196y1994}. 
Both B2 austenites (FeRh below 353~K and NiTi above 313~K) have a premartensitic instability with known unstable phonon modes  \cite{PRB90p060102y2014,PRL113p265701y2014}. 
While they both show elasto- and baro-caloric responses, FeRh also exhibits a giant magnetocaloric effect at its metamagnetic transition from an antiferromagnetic (AFM) to a ferromagnetic (FM) state at the critical temperature $T_c$ of $353\,$K, with a $1$\%  decrease in density \cite{PSSB20n1pK25y1967}. 
Properties of FeRh were extensively studied 
experimentally  \cite{JAP37n3p1257y1966,JAP38p1263y1967,PSSB20n1pK25y1967,Nikitin1990p363,Annaorazov1996,Fallot1938,Fallot1939,Bergevin1961,JAP33n3p1343y1962,PhysRev131p183y1963,JETP19n6p1348y1964,PhysRev134pA1547y1964,JAP35n3p938y1964,JETP23n6p984y1966,JPhysC3n1SpS46y1970,Vinokurova1981,Annaorazov1992,Intermetallics15n9p1237y2007,IEEEtransMag44n11p2875y2008,PRB81p104415y2010,PRL109p255901y2012,PRB92p184408y2015,jjimm80n3p186y2016,ActaMat106p15y2016,SciRep6p22383y2016,JPhysD51p024003y2018Loving,SciRep8n1p1778y2018,JMMM459p182y2018} 
and theoretically     \cite{JETP36n1p105y1973,SovPhysUsp11n5p727y1969,Hasegawa1987p175,PRB46p2864y1992,PRB46p14198y1992,JAP90n12p6251y2001,PRB83p174408y2011,PRB89p054427y2014,PRB91p014435y2015,PRB91p224421y2015,PRB92p094402y2015,PRB93p024423y2016,PRB94p014109y2016,PRB94p174435y2016,2016.PRB.94.180407,PRB97p140407y2018}.  
Notably, the  metamagnetic  $T_c$ of FeRh is sensitive to stoichiometry, lowering precipitously with small additions of at.\%Rh \cite{PhysRev131p183y1963}. 
A giant caloric effect  is found at this transition in the quenched Fe$_{49}$Rh$_{51}$ sample \cite{Nikitin1990p363}, i.e., a directly measured temperature drop of $12.9\,$K at $1.95\,$Tesla.

{\par} 
While bulk FeRh is prohibitively expensive, 
Fe--Rh may find use in caloric thin-film
\cite{Nikitin1990p363,ActaMat106p15y2016,Annaorazov1992,JPhysD51p024003y2018Loving,JPhysD41n19p192004y2008,JAP103n7p07F501y2008,ScriptaMat67n6p566y2012,PRB89p214105y2014,NComms11614y2016,JPhysD51n10p105001y2018,MRL6p106y2018,PhysRevApplied9p034030y2018,ApplSurSci449p380y2018} 
 and nano\-scale devices   \cite{SciRep6p22383y2016,PRB94p174435y2016,APL82n17p2859y2003,PRL93p197403y2004,IEEE40n4p2537y2004,PRB82p184418y2010,NMat13p345y2014,NMat13n4p367y2014,PRL116p097203y2016,AIPadvances6n1p015211y2016,ACSCatalysis8p7279y2018}.  
Nonetheless, and notably here, it mainly serves as a well-studied but suitably complex system to test methods for reliability in thermodynamic assessments and prediction of caloric properties, specifically because it exhibits instabilities from anharmonic atomic motion, which affects caloric behavior.
The FeRh groundstate and a martensitic transformation in the AFM phase at cryogenic temperatures were recently addressed \cite{2017.FeRh.gs}.

{\par} Here we focus on estimators \cite{JPhysD51n2p024002y2018} to predict thermodynamics at the meta\-magnetic transformation near room temperature.  
We find that quantities relevant to calorics can be calculated in a quantitative agreement with measurements (Table~\ref{t3}).   
We also provide insight into the key requirements to predict caloric behavior accurately -- necessary to identify the computational screening measures and correlations that assist in materials discovery  \cite{Complexity11p36y2006}.   
While some computations can be intensive (e.g., phonons and lattice entropy), 
the results are useful for testing faster estimators  \cite{Kubler2009,Vonsovsky1971,PRB90p174107}.

\begin{table}[t]
\centering
\begin{tabular}{lllll}
\hline
         & Theory & Expt. & Units & Ref. \\
\hline
$M$ (FM) & 149 & $\approx \! 150$ & $\mbox{A}\,\mbox{m}^2 \, \mbox{kg}^{-1}$ & \cite{jjimm80n3p186y2016} \\  
$a$ (FM) & 3.012 & 2.997  & \AA & \cite{PSSB20n1pK25y1967,JETP19n6p1348y1964} \\
$a$ (AF) & 2.996 & 2.987  & \AA  & \cite{PSSB20n1pK25y1967,JETP19n6p1348y1964} \\
$\Delta S_T (T_c)$ & 11.9 & 12--14  & $\mbox{J}\,\mbox{kg}^{-1}\mbox{K}^{-1}$ & \cite{ActaMat106p15y2016,PRB89p214105y2014,PhysLettA171p234y1992,Richardson1973,JAP37n3p1257y1966} \\
$\Delta S $ &   & 17--19  & $\mbox{J}\,\mbox{kg}^{-1}\mbox{K}^{-1}$ & \cite{PRL109p255901y2012,JETP36n1p105y1973,JETP19n6p1348y1964} \\
$-\Delta T_S$ & 13 & 10--13  & K & \cite{ActaMat106p15y2016,PRB89p214105y2014} \\
$T_c$ (AF--FM) & 346 & 353  & K & \cite{PhysRev131p183y1963} \\
\hline
\end{tabular}
\caption{\label{t3} Calculated (Theory) and experimental (Expt.) magnetization, lattice constant, caloric effect, and phase transition temperature. References 
are given for experiment.  
$\Delta S $ differs from $\Delta S_T (T_c)$ 
due to incorrect assessment, \cite{JETP19n6p1348y1964,PRL109p255901y2012}  see sections~\ref{a5} and \ref{Discussion}. 
}
\end{table}

{\par}  Computational details are provided in section~\ref{ComputationalDetails}.
In section~\ref{Results}, we address the caloric effects and calculate $\Delta S_T$ and $\Delta T_{S}$ at the meta\-magnetic transition.  
Importantly, in subsection \ref{a5L} we test a method for addressing non-harmonic atomic vibrations at a relevant temperature, because the commonly-used linear-response and small-displacement methods employed to assess lattice  entropy fail near lattice instabilities, including those that arise from anharmonicity of the atomic potential energy surface.
In section~\ref{Estimates}, we offer estimators of enthalpy change $\Delta H$,  transition temperature $T_c$, and its derivative $dT_c /dB$ with respect to the external field $B$. 
Some of the issues and limitations of the common and alternative approaches are discussed in section~\ref{Discussion}.  
Generic remarks about the upper bounds, chemical disorder, and hysteresis are provided in section~\ref{Discussion2}.
Finally, in section~\ref{Beyond}, screening for better caloric systems beyond ReRh, as now verified experimentally, is presented, followed by a summary in section~\ref{Summary}. 
Thus, we review and assess the relevant methods and estimates of caloric properties, as showcased in a test system (FeRh), but which may be applied quite generally.

\section{\label{ComputationalDetails}Computational methods}
For FeRh compound, density functional theory (DFT) calculations were performed using the Vienna \emph{ab initio} simulation package ({\tt VASP}) \cite{VASP1,VASP2}. 
We used projector augmented waves (PAW) \cite{PAW,PAW2} and the PBE exchange-correlation functional \cite{PBE} with Vosko-Wilk-Nusair spin-polarization  \cite{VOSKOWN}, combined with a modified Broyden method \cite{PRB38p12807y1988} for accelerated convergence. 
Brillouin zone integrations were performed on a Monkhorst-Pack mesh \cite{MonkhorstPack1976} with $\ge 50$ $k$-points per {\AA}$^{-1}$ with $\Gamma$ included. 
The plane-wave basis-set energy cutoff  was increased to 334.9 eV (or 511.4 eV for  augmentation charges) by the high-precision flag. 
During computing of the atomic forces, an additional (third) support grid was used for the evaluation of the augmentation charges.   

In non-stoichiometric cases, chemical disorder was addressed using either super\-cells or the coherent potential approximation (CPA) \cite{CPA2}, 
implemented in the KKR code {\tt MECCA} \cite{MECCA}. 
Components of the {\tt TTK} toolkit \cite{PRL92p255702y2004} were used to prepare the super\-cells.

{\par}As needed for barriers or saddle-point transitions \cite{2017.FeRh.gs}, DFT was combined with a generalized solid-state nudged elastic band (GSS-NEB) \cite{SSNEB}, which includes a built-in 
{\tt C2NEB} algorithm \cite{C2NEBsoft} with two climbing images \cite{C2NEB}. 

Phonons were calculated using the finite atomic displacement method, implemented in the {\tt Phon} code \cite{Phon}. The force-constant matrix  \cite{Phon} was constructed from the atomic forces (in the file FORCES), computed using {\tt VASP}.  
The atomic displacements varied from 0.04 to 0.12 {\AA} in a cubic $4 \!\times\! 4 \!\times\! 4$ supercell containing 64 FeRh formula units (f.u.). 
The phonon density of states was computed \cite{Phon} using $21^3$ $k$-point grid in the reciprocal space (LRECIP=.TRUE.) and the Gaussian smearing (DOSSMEAR=$0.05\,$THz); 
the output file THERMO provided the lattice entropy $S_L$ at finite $T$. 
We used the experimental atomic masses: $m(\mbox{Fe})= 55.845$ and  $m(\mbox{Rh})= 102.9055$ atomic unified mass units, u=$\frac{1}{12}m(^{12}\mbox{C})$.
We also present a method that more properly addresses anharmonic vibration near instabilities, which has a significant affect on entropy.

\section{\label{Results}Results} 
{\par} The magneto\-structural transition in B2 FeRh between FM and AFM phases (Fig.~\ref{fig1}) is accompanied by a change of electronic structure (Fig.~\ref{figDOS}), energy and volume (Fig.~\ref{figEPV}). While an electronic transition happens with the speed of light, a structural transformation (including volume change) propagates no faster than the speed of sound  \cite{PRB91p174104}.  
So, the electronic transformation is accompanied by discontinuity in pressure that drives the volume change \cite{JChemPhys142p064707y2015}. 
The possible causes for electronic transitions include initial structure change or application of an external field.  In particular, as is well established, the magneto\-structural transformation of FeRh can be caused by application of an external magnetic field and/or stress, strain, or thermal expansion.

\subsection{\label{atomicM} Spin Density and Itinerant Magnetism}
\begin{figure}[h]
\centering
\includegraphics[width=86mm]{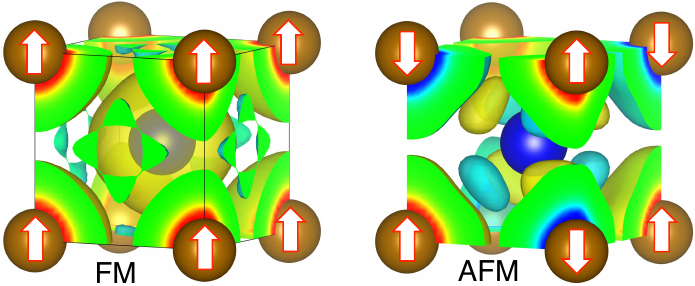}  
\caption{\label{fig1}  (Color online).  
In B2 FeRh, the electronic spin density as $\pm 0.002~e$/{\AA} isosurfaces in (001) FM (left) and (111) AFM (right) spin configurations.}
\end{figure}

\begin{figure*}
\centering
\begin{minipage}[b]{0.47\textwidth}
\includegraphics[width=56mm]{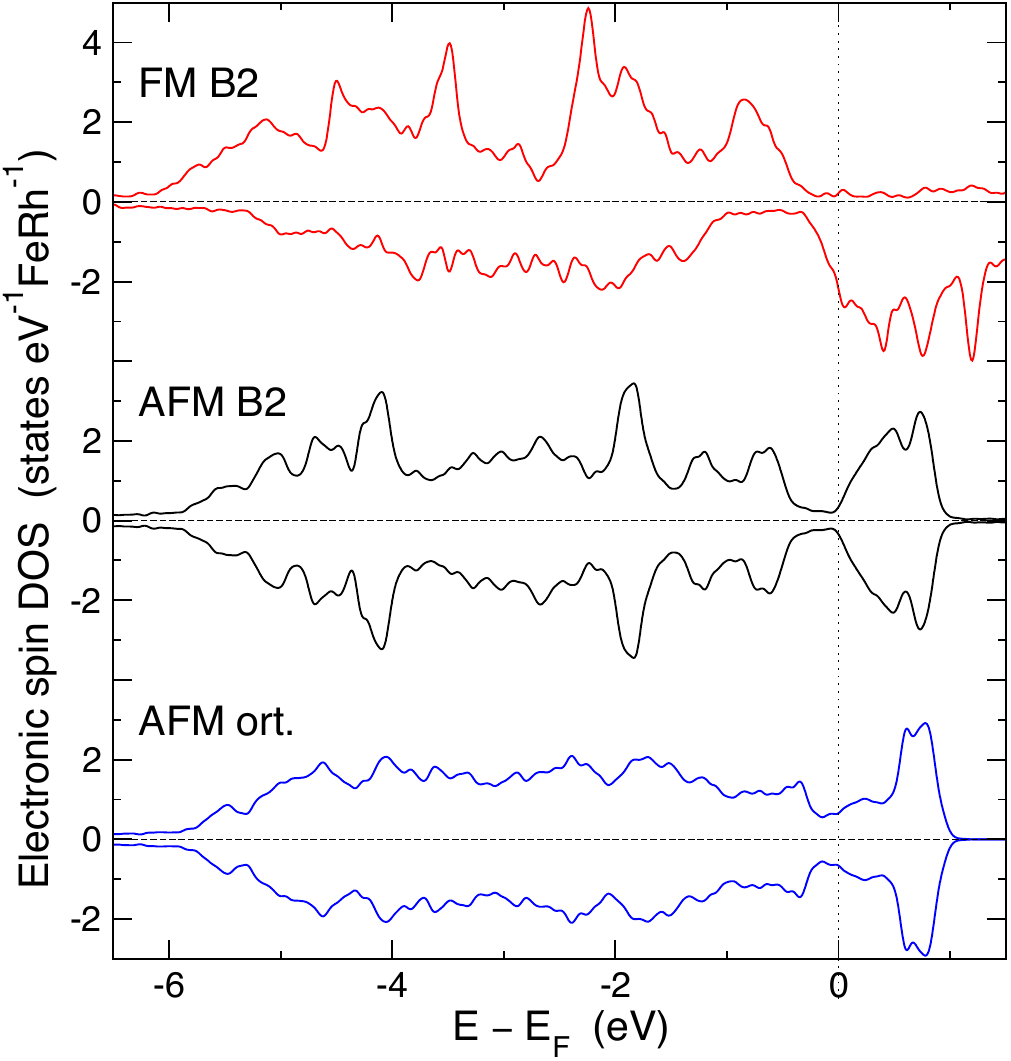}  
\caption{\label{figDOS} (Color online). For B2 FeRh, electronic DOS for FM (upper) and AFM spin ordering in ideal cube, 
and AFM orthorhombic structure (lower) at zero pressure \cite{PRB97p140407y2018}. }
\end{minipage} \hfill
\begin{minipage}[b]{0.47\textwidth}
\includegraphics[width=60mm]{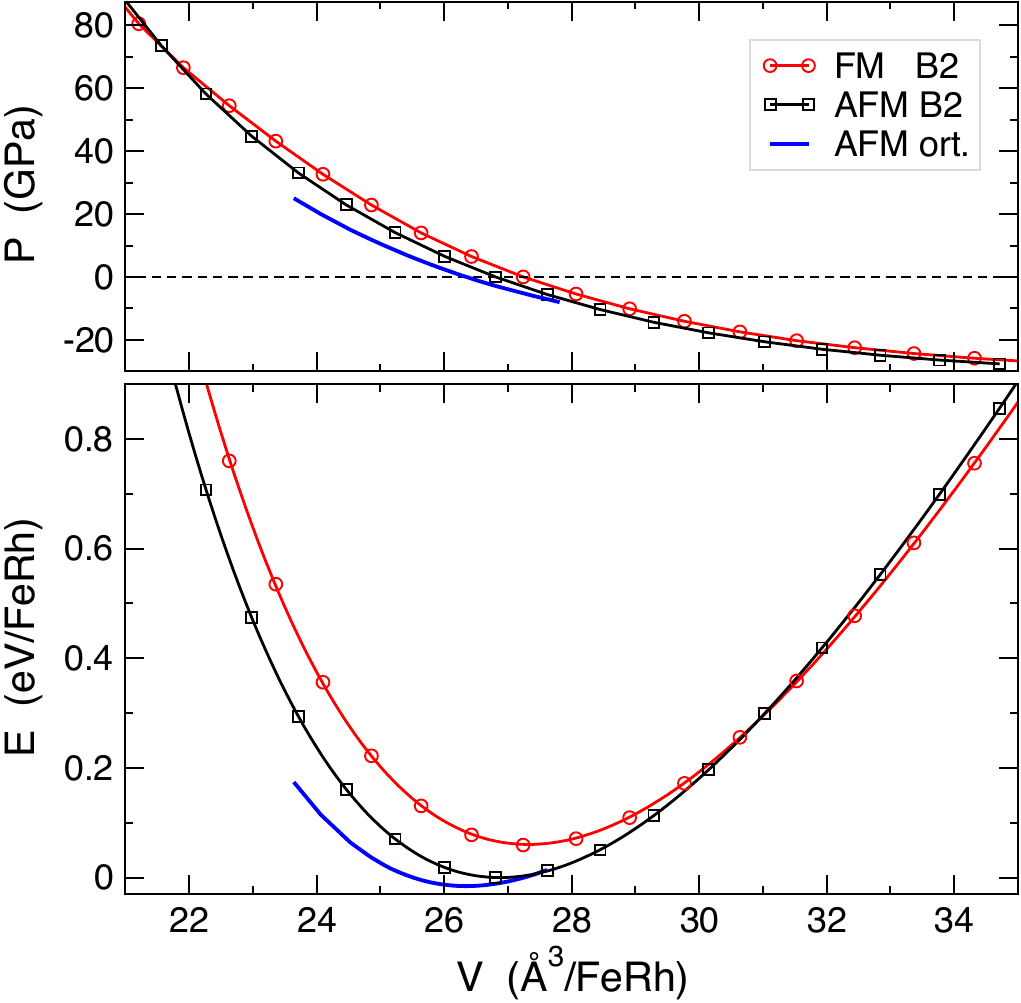}   
\caption{\label{figEPV} (Color online). Pressure $P$ (GPa) and energy $E$ (eV)   vs. volume $V$ ({\AA}) per FeRh formula unit for B2 FM and AFM, and orthorhombic AFM structures. }
\end{minipage} \hfill
\end{figure*}

{\par} Figure~\ref{fig1} shows the real-space distribution of the electronic spin density in the B2 cubic cell of FeRh, which is an itinerant magnet.  
Importantly, spin density around Rh atoms is not zero in both phases, 
but the atomic magnetic moment of Rh is zero in an ideal B2 AFM structure due to the inversion symmetry with a center at Rh nucleus.  Indeed, if the distribution of Fe moments is symmetric in the AFM phase, then the electronic spin density sums to zero within the Rh atomic sphere (and within an arbitrary Rh-centered sphere of any radius).  However, any asymmetry due to the fluctuating Fe-Rh distances or Fe moments (e.g., due to thermal disorder)  would result in a non-zero atomic magnetic moment of Rh. 

{\par} At the AFM-FM phase transition, the calculated magnetization changes from zero to 149 A$\,$m$^2$/kg (4.2~$\mu_B$/FeRh). 
With caution, one can integrate the spin density inside each atomic sphere to find the ``atomic'' magnetic moment.  
We find that the Rh moments change from 0 (AFM) to 1 $\mu_B$ (FM), and the Fe moments change from $\pm$ 3.1 (AFM) to 3.2 $\mu_B$ (FM).

\subsection{\label{a5e}Electronic and Magnetic Entropy}  
{\par} As seen in Fig.~\ref{figDOS}, the total electronic spin density of states (DOS) at the Fermi energy $E_F$ changes substantially during the  transformation  from $n(E_F) = 0.677$ in the AFM to $2.310$ states/eV per FeRh formula unit (f.u.) in the FM state. Contributions of both spins are equal in the AFM, while minority spins dominate at $E_F$ in the FM state (Fig.~\ref{figDOS}). 

{\par} The electronic entropy (as estimated by the Sommerfeld's expansion) is  
\begin{equation}
\label{eSommerfeld}
S_e (T) \approx (\pi^2/3) \cdot k_B^2 T \cdot n({E}_F) , 
\end{equation}
which yields $0.23$ (FM)  and $0.07$ (AFM) $\, k_B$/FeRh at $T_c$.  
The difference $\Delta S_e$ is $0.163\, k_B$/FeRh (or $0.08 \, k_B$ per atom). 
The Sommerfeld approximation in most cases tested has been a reasonably reliable approximation between structural variants arising at solid-solid phase transitions.

{\par} Spin-polarized electrons are responsible for both conductivity and magnetism; they account for both electronic and magnetic contributions to the entropy, as required in an itinerant magnet \cite{Kubler2009}, such as FeRh. 
Fluctuations of atomic magnetic moments can be expanded in an electronic basis in both FM and AFM phases. 
$S_e$ includes entropy of thermal excitations in both spin channels (i.e., electronic and magnetic contributions). 

{\par} The total entropy is $S = \ln \Omega$, where $\Omega$ is the number of accessible micro\-states in the whole system. 
Typically, magnetic entropy is small in the FM and AFM states, where the number of magnetic states (per atom) is close to 1, 
and it is larger in a paramagnetic (PM) state, which is not relevant to the AFM--FM phase transition. 
In decomposing the total entropy into electronic, magnetic, and lattice contributions, 
sometimes mistakes were made \cite{PRL109p255901y2012},  
leading to notably wrong findings. 
\footnote{
In Ref.~\onlinecite{PRL109p255901y2012},
the total integrated entropy difference 
($\Delta S = 17 \pm 3$~J/kg/K)
was incorrectly decomposed into 
lattice 
($\Delta S_{latt} = -33 \pm 9$~J/kg/K), 
electronic
($\Delta S_{el} = 8 \pm 1$~J/kg/K), 
and magnetic 
($\Delta S_{mag} = 43 \pm 9$~J/kg/K) 
contributions. 
The negative sign of the lattice contribution is notable.
} 
We discuss the  
issues with indirect assessments in section~\ref{Discussion}.

\subsection{\label{a5V}Compression and Expansion}
The energy $E$ and pressure $P$ versus volume $V$ curves for the main competing structures in FeRh are shown in Fig.~\ref{figEPV}. 
The FM B2 and AFM B2 are the terminal states of the metamagnetic phase transition, accompanied by the magneto\-caloric effect; 
the AFM orthorhombic (martensitic) ground state of FeRh is discussed elsewhere \cite{2017.FeRh.gs}.
The calculated equilibrium lattice constants $a=V^{1/3}$ are compared to experiment in Table~\ref{t3}, less than +0.2\% difference from experiment using a PBE density functional.
The FM and AFM states have a crossover at higher volumes. 
From these plots, the meta\-magnetic transition already can be anticipated. 
At zero pressure $P$, 
 the FM state is  $\delta H_0 = 29.8 \pm 1.0 $~meV/atom above the AFM state   
($\delta H_0 / k_B = 346 \pm 12 \,$K, 
i.e., near the measured $T_c = 353 \,$K, see section~\ref{Estimates}).

In addition,  a pre\-martensitic instability is anticipated in B2 AFM state with known phonon instabilities \cite{PRB93p024423y2016,PRB94p014109y2016,PRB94p174435y2016}, 
and a martensitic transformation 
from B2 AFM austenite to orthorhombic AFM martensite 
at cryogenic $T$
was suggested  
by direct GSS-NEB calculations \cite{PRB97p140407y2018}.

\subsection{\label{a5L}Lattice Entropy -- Anharmonic and Harmonic Vibrations}  
{\par} Vibrational entropy in materials can contribute significantly to their caloric response.  
To assess vibrational entropy of phonon excitations at a finite $T$, 
the standard approach is to calculate the quasi\-harmonic phonon frequencies 
 by linear-response or small atomic displacement method. 
However, both of these methods inherently assume a harmonic atomic potential. 
In materials with structural and magnetic instabilities (or, more generally, ``dimpled'' potential energy surfaces), this assumption is invalid,  
at least near temperatures, where the crossover between states occur and key associated properties manifest. 
With the pre\-martensitic instability \cite{2017.FeRh.gs} in AFM(111) B2 FeRh, 
similar (but smaller) to that in NiTi austenite \cite{PRB90p060102y2014,PRL113p265701y2014}, 
care must be taken to calculate accurately the lattice entropy.

\begin{figure*}
\centering
\begin{minipage}[b]{0.45\textwidth}
\includegraphics[width=60mm]{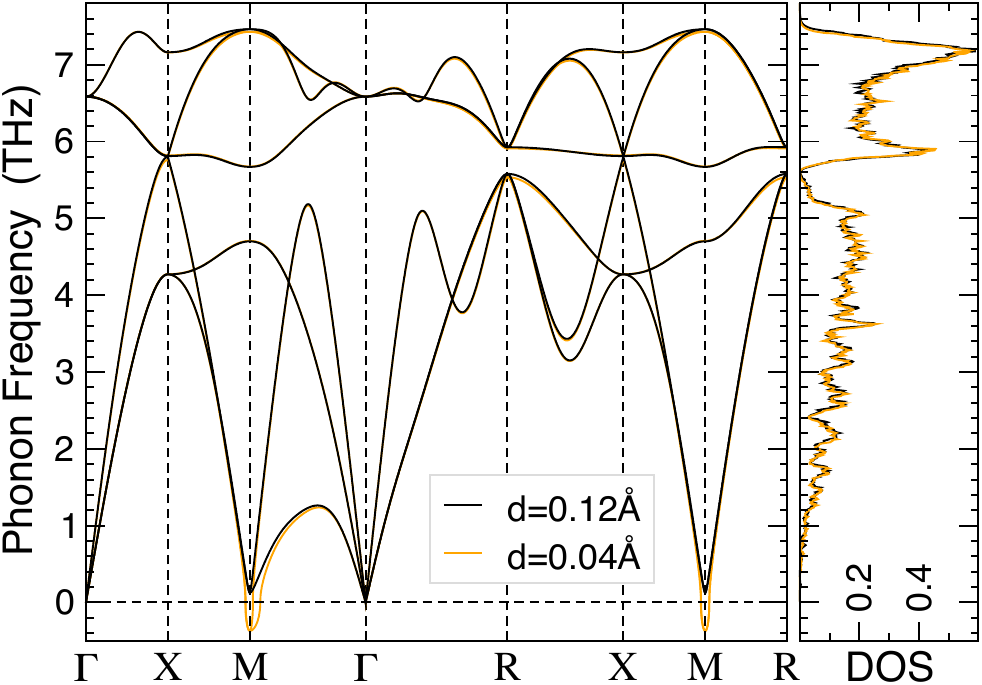}  
\caption{\label{figPhonAFM}  (Color online).  Phonon frequencies and DOS for AFM B2-FeRh ($2.996\,${\AA}) using small (0.04\AA) and large (0.12\AA) displacements. 
Unstable phonons at $M$ ($\frac{1}{2} \frac{1}{2} 0$) appear at $d \le 0.06\,${\AA}.  }
\end{minipage} \hfill 
\begin{minipage}[b]{0.47\textwidth}
\centering 
\includegraphics[width=60mm]{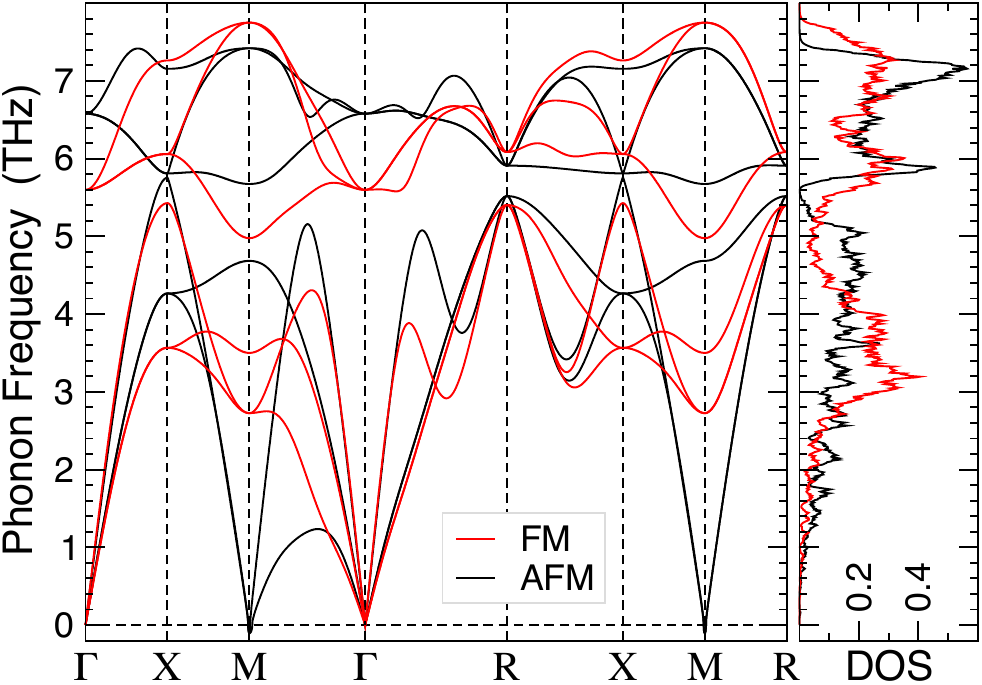}  
\caption{\label{figPhon}  (Color online).  Phonon frequencies and DOS for FM and AFM states in B2-FeRh ($2.996\,${\AA}) evaluated with $d(\mbox{Fe}) \! \approx \! d(\mbox{Rh}) \! \approx \! 0.06\,${\AA} at $\frac{1}{2}k_B T_c$, see text  and  Fig.~\ref{figEd}, now no unstable phonons at $M$ ($\frac{1}{2} \frac{1}{2} 0$).}
\end{minipage} \hfill  
\end{figure*}

{\par} Here, distinct from previous work, we evaluate phonon frequencies and density of states (DOS) along with their sensitivity to the atomic displacement $d$ used to calculate them.  Using the small-displacement method \cite{Phon} at zero pressure, we find an unstable  phonon mode at $M$ $(\frac{1}{2} \frac{1}{2} 0)$ in the  AFM state (Fig.~\ref{figPhonAFM}), but not in the FM state (Fig.~\ref{figPhon}), 
as also found in recent publications \cite{PRB94p014109y2016,PRB94p174435y2016,2016.PRB.94.180407}. 
At ambient conditions, FM FeRh is stable but not harmonic, with instabilities nearby (e.g., due to strain) \cite{PRB94p014109y2016}. 
For the least harmonic phonons, frequency dependence on $d$ is the strongest. 
The high-frequency optical phonon modes are harmonic in FM and AFM phases, while the low-frequency acoustic phonons show less harmonic behavior around $M$, where the difference between modes, calculated for $d=0.04$ and 0.12~{\AA}, is the largest (Fig.~\ref{figPhonAFM}).

{\par}Notably, the M-point phonon instability leads to a cryogenic martensitic transition in AFM FeRh with atomic shuffles of $d_{Fe}=0.061$ and $d_{Rh}=0.053$ in fractional lattice coordinates, showing that atomic potentials have dimples around the high-temperature symmetric structure (B2) 
and are inherently anharmonic. \cite{PRB97p140407y2018}  
One anticipates then a $d$-dependence of phonon frequencies, which are well-defined at each fixed $d$.

\begin{figure}
\centering
\includegraphics[width=70mm]{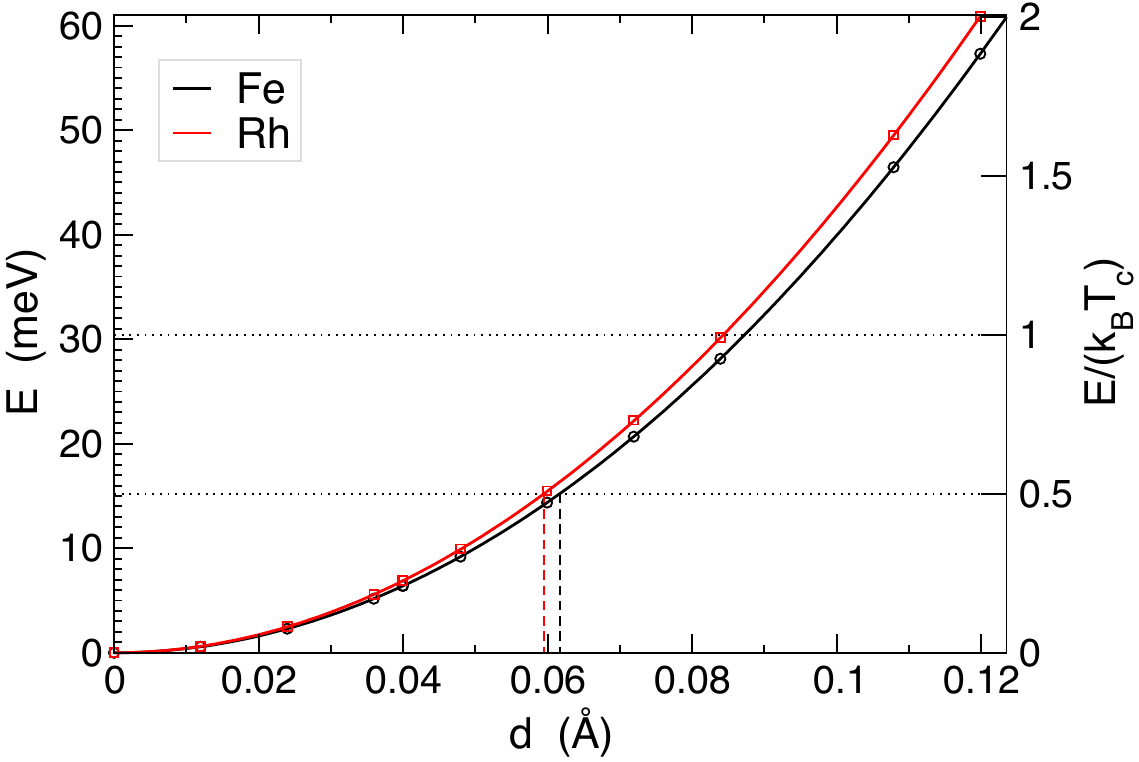}  
\caption{\label{figEd} (Color online). For FM B2 FeRh ($2.997\,${\AA}), energy (E, meV) versus displacement $d$ of a Fe (black) or Rh atom (red)  along [100] compared to $k_B T_c$. 
Lines are quartic fits.}
\end{figure}

{\par}  To calculate phonons at a given temperature $T$,  
one could use thermal atomic displacements and forces from  \emph{ab initio} molecular dynamics (MD), 
say, in the {\tt ThermoPhonon} code \cite{PRB90p060102y2014,ThermoPhonon}. 
A faster, albeit more approximate, method (which we use at $T_c$ of $353\,K$)  
is an application of the quasi\-harmonic approximation with a finite single-atom displacement $d$ scaled to a ``thermal'' potential energy $E(d) = \frac{1}{2}k_B T$ in an ideal structure  (Fig.~\ref{figEd}). 
This method is applied to FeRh in Fig.~\ref{figPhon} and shows  that AFM B2 structure has an unstable phonon mode at $M$ with an amplitude of only $0.1\,i$ THz (i.e., close to zero) at  ``thermal'' displacements $d (T_c) \approx 0.06\,${\AA} [here $d(\mbox{Fe}) \ne d(\mbox{Rh})$, see Fig.~\ref{figEd}];  this instability becomes larger at smaller $d$ (including infinitesimal case used in linear-response methods) and disappears at larger $d$. 

{\par} As phonons in both AFM and FM phases are anharmonic,
and the lattice entropy $S_L$ is affected mostly by the soft phonon modes, 
this finite-displacement method within a quasi\-harmonic approximation \cite{Togo2015} is 
a reliable computational ``trick'' to avoid unstable phonons at a relevant finite temperature;   
it uses substantially less computational time than the other method based on MD at fixed $T$ \cite{PRB90p060102y2014}, 
 while yielding correct estimates.
 
\begin{figure*}[h]
\centering
\begin{minipage}[b]{0.45\textwidth}
\includegraphics[width=58mm]{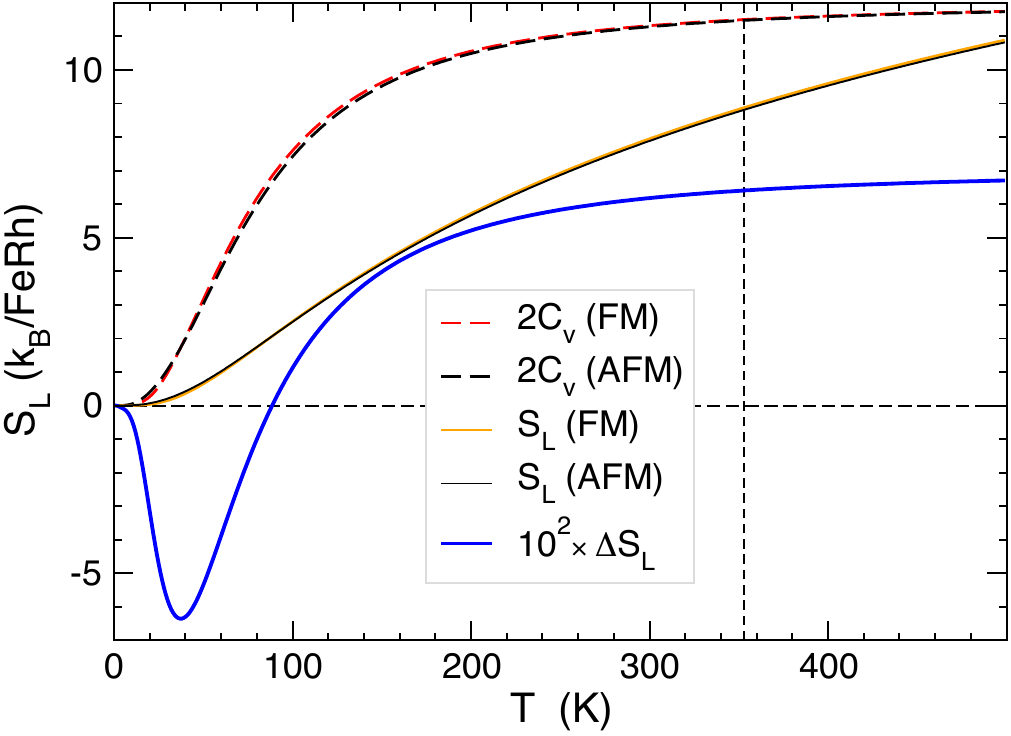} 
\caption{\label{figCv} (Color online). For B2 FeRh, heat capacity $C_V$,  $S_L$, and  $\Delta S_L $(FM$-$AFM) from phonons (Fig.~\ref{figPhon}) using $d(T_c) \approx 0.06\,${\AA} from Fig.~\ref{figEd}.  T$_c$ at $353\,$K is denoted by vertical dashed line.}
\end{minipage} \hfill 
\begin{minipage}[b]{0.45\textwidth}
\centering
\includegraphics[width=58mm]{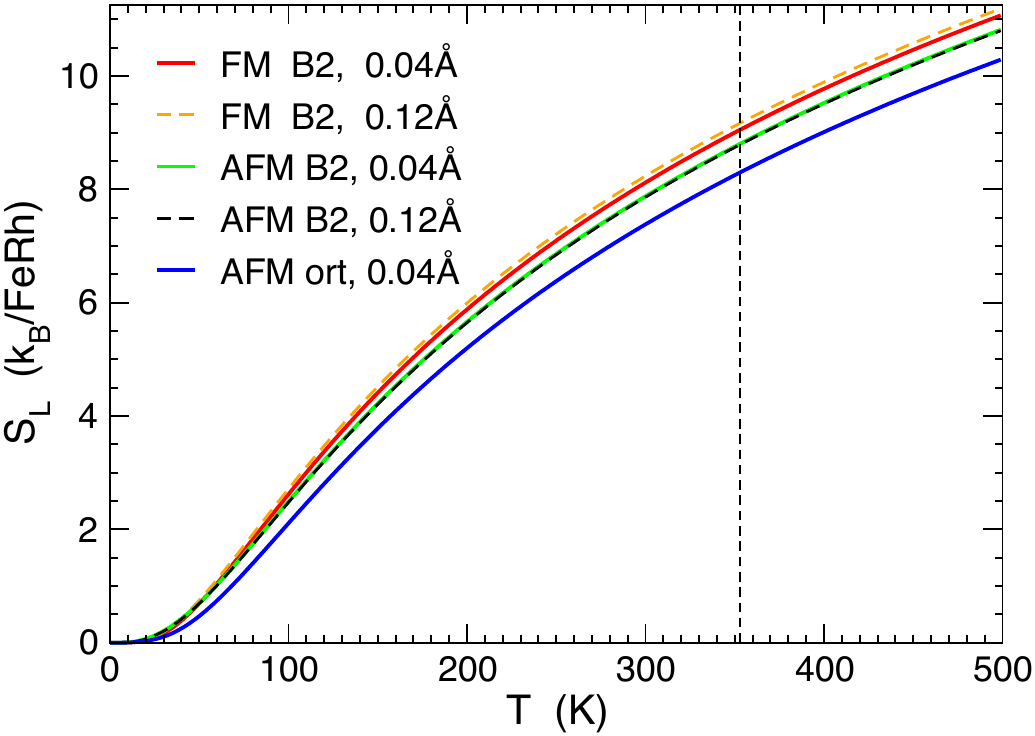} 
\caption{\label{figS} (Color online). For B2 and orthorhombic FeRh structures in FM and AFM states at $P$=0, lattice entropy $S_L$ using small (0.04 \AA) and large (0.12 \AA) atomic displacements $d$.  }
\end{minipage} \hfill 
\end{figure*}

 {\par} The atomic displacement $d (T)$ can be adjusted to temperature $T$ (Fig.~\ref{figEd}) 
and used to evaluate the $T$-dependent lattice entropy $S_L [T, d(T)]$, calculated  at fixed lattice constants.
Below we use the phonon DOS to evaluate $\Delta S_L$ at the metamagnetic transition at $T_c$ (Fig.~\ref{figCv}).
Importantly,  due to anharmonicity and finite thermal displacements at finite temperature, 
$\Delta S_L [T_c, d(T_c)]$ is increased by 50\%,  
compared to $\Delta S_L [T_c, d \to 0]$.

{\par}   In particular, for FM B2 FeRh, the energy $E$ versus atomic displacement $d$ (shown in Fig.~\ref{figEd}) 
can be fit well by a quartic (not quadratic) polynomial, i.e., $E(d)=E^{(2)} d^2 + E^{(4)} d^4$. 
We find $E^{(2)}_{Fe} = 4.003\, \mbox{eV/\AA}^2$ and $E^{(4)}_{Fe} = -1.030 \, \mbox{eV/\AA}^4$ for Fe and  $E^{(2)}_{Rh} = 4.317\, \mbox{eV/\AA}^2$ and $E^{(4)}_{Rh} = -5.496 \, \mbox{eV/\AA}^4$ for Rh.
Consequently, $S_L$  depends on the atomic displacement $d$, see Fig.~\ref{figS}. 
In the FM phase, it  changes from $8.859\, k_B$/FeRh for small $d\!=\!0.04 \,${\AA} to $8.972\, k_B$/FeRh for large $d\!=\!0.12 \,${\AA} at fixed $a = 2.997\,${\AA}.
In the AFM phase, we find a small change from 8.8176 to $8.7945\, k_B$/FeRh for the same fixed values of $d$, see Fig.~\ref{figS}. 
Interestingly, the unstable AFM B2 phase is less anharmonic than the stable FM B2 phase, 
which develops phonon instability at a strain \cite{PRB94p014109y2016}. 
The small-$d$ ($0.04 \,${\AA}) method  provides $\Delta S_L$ of  $0.042 \, k_B$/FeRh (or $0.02 \, k_B$ per atom).
However, with $T$-dependent displacements $d(T_c) \approx 0.06\,${\AA} ($d_{Fe} \ne d_{Rh}$, see Fig.~\ref{figEd}),   
$\Delta S_L [T_c, d(T_c)]$ increases by 50\% 
to  $0.064 \, k_B$/FeRh (or $0.03 \, k_B$ per atom), see Fig.~\ref{figCv}.
Thus, for FeRh, the spin-polarized electrons, fully accounted here,  
provide the leading contribution to the total entropy change $\Delta S_T (T_c)$, 
while the lattice entropy contribution is smaller (only 28\%), 
but not negligible.
This relative contribution agrees with an early prediction \cite{JETP36n1p105y1973} and its recent confirmation \cite{PRB89p054427y2014}. 
Nonetheless, for FeRh,  $\Delta S_L (T_c)$, now increased by 50\% 
from anharmonicity, is $\approx 40$\% 
of the calculated electronic contribution 
$\Delta S_e (T_c)  = 0.163\, k_B$/FeRh. 

{\par}Anharmonicity affects the phonons and associated thermodynamic quantities. 
In general,  anharmonic effects must be properly included in a consideration of thermodynamics near lattice instabilities and phase transitions. 
Here, we have described a quick method to include these $T$-dependent effects in anharmonic systems by probing the amplitude of atomic displacements $d$ dependence of the vibrational frequencies.
 If the phonons were harmonic, then the lattice entropy $S_L [T, d(T)]$ would not depend on $d$  \cite{Phon,Togo2015}; 
 Fig.~\ref{figS} shows that in FeRh this is not the case, as $d$ in the FM B2 state is larger than for the AFM state.

\subsection{\label{a5}Entropy Change}  
{\par}The total entropy includes the electronic (with magnetic) and lattice contributions.
We calculate the total entropy change $\Delta S = S(\mbox{FM}) - S(\mbox{AFM}) $ due to electronic transformation  at $T_c = 353\,$K at fixed lattice constant $a = 2.997\,${\AA} (measured \cite{PSSB20n1pK25y1967} in the Fe$_{50}$Rh$_{50}$ FM phase at $T_c $). 
We find $\Delta S_T = 0.227 \, k_B$/FeRh, or $0.11\,k_B$/atom  (i.e., $11.9 \,$J$\,$kg$^{-1}$K$^{-1}$) 
for the isothermal total entropy change at the  meta\-magnetic transformation at $T_c$ at fixed volume.  
The lattice entropy contribution $\Delta S_L$ is  28\% of $\Delta S_T$; 
ignoring the anharmonic effects would lead to a 50\% relative error in $\Delta S_L$ and 14\% error in $\Delta S_T$. 

{\par} In experiment,  the maximum total entropy change of $12.5 \pm 1$~J$\,$kg$^{-1}$K$^{-1}$ was the same for both baro- and magneto-caloric effects  in Fe$_{49}$Rh$_{51}$ \cite{PRB89p214105y2014}. 
Three assessment methods 
gave comparable values for $\Delta S_T$ for the Fe$_{49}$Rh$_{51}$ \cite{ActaMat106p15y2016},  namely, 
calorimetry: $12.1$,  Clausius-Clapeyron: $13.1$, and Maxwell relations: $13.6$~J$\,$kg$^{-1}$K$^{-1}$. 
An earlier measurement \cite{JAP37n3p1257y1966} yielded $14.0~$J$\,$kg$^{-1}$K$^{-1}$ for stoichiometric FeRh  
and found a compositional dependence of  $\Delta S$ for the samples doped with Pd, Pt, or Ir.  
The experimental values [in J$\cdot$kg$^{-1}$K$^{-1}$] for $\Delta S_T$  of 
13.6 \cite{ActaMat106p15y2016},
$12.5 \pm 1$ \cite{PRB89p214105y2014}, 
$13 \pm 1$ \cite{PhysLettA171p234y1992},
$13 \pm 1$ \cite{Richardson1973}, and 
14 \cite{JAP37n3p1257y1966} 
differ from the higher assessed values of
$19 \pm 2$  \cite{JETP19n6p1348y1964}, 
18.3 \cite{JETP36n1p105y1973}, and
$17 \pm 3$~J$\,$kg$^{-1}$K$^{-1}$  \cite{PRL109p255901y2012};   
inaccuracies in Refs.~\onlinecite{JETP19n6p1348y1964,JETP36n1p105y1973,PRL109p255901y2012} originated from
 subtracting values measured at two different compositions \cite{PRL109p255901y2012},  
using \cite{JETP19n6p1348y1964} $\Delta V_P$ instead of $\Delta V_T$  in the Clausius-Clapeyron equation (\ref{ClausiusClapeyron}), 
see section~\ref{Discussion},
or increasing the extrapolated value of  $(d B_c / d T)_{B=0}$ in eq.~\ref{dTdB}  
to account for an overestimated 
magneto\-caloric effect in a Fe$_{0.48}$Rh$_{0.52}$ sample \cite{JETP36n1p105y1973}. 
The assessed values depend on the method \cite{ActaMat106p15y2016}, sample composition \cite{JAP37n3p1257y1966}, and preparation \cite{Nikitin1990p363}. 
The calculated and experimental values are compared in Table~\ref{t3}.

\subsection{\label{a6}The Caloric Effect}  
The maximum isentropic temperature change is 
\begin{equation} 
\label{dTs}
 \Delta T_{S}  =  -T_c \Delta S_T / C_B . 
\end{equation}
Here $C_B (B,T)$  is the heat capacity at constant magnetic field $B$.   
Using the asymptotic limit  $C_B$\,$\approx$\,$3 k_B$/atom ($6 \, k_B$/FeRh or 314~J$\cdot$kg$^{-1}$K$^{-1}$) for solid FeRh at $T \ge 300\,$K (Fig.~\ref{figCv})  
and our value of $\Delta S_T$ (section~\ref{a5}), we find 
$\Delta T_{S} = -13\,$K. 
This value agrees  with the experimental assessments \cite{ActaMat106p15y2016}, ranging  from $-10.6$ to $-12\,$K, see Table~\ref{t3}. 
However, it differs from an early estimate \cite{JETP36n1p105y1973} of 
$ \Delta T_S =  - (20 \pm 2)\,$K, obtained using too high value of $\Delta S = 18.3 \,$J$\,$kg$^{-1}$K$^{-1}$ in  eq.~\ref{dTs}. 
The directly measured  adiabatic temperature change $\Delta T_{ad} (\Delta B)$, produced by  an added external field of $\Delta B \! =  \! 1.95 \,$Tesla, can be as large as  $-12.9\,$K for the quenched Fe$_{49}$Rh$_{51}$ samples \cite{Nikitin1990p363}.

\section{\label{Estimates}Estimators for Materials Screening}

\subsection*{\label{EstimateDE}Isothermal enthalpy change $\Delta H_T (T_c)$}
From Gibbs relation, the isothermal enthalpy change $\Delta H_T$ at $T_c$  is the key quantity, given by the formally exact equation
\begin{equation}
\label{eDE}
  \Delta H_T (T_c) = T_c \cdot \Delta S_T (T_c). 
\end{equation}
Using either experimental or calculated (below) $T_c$ and calculated $\Delta S_T (T_c) = 11.9 \,$J$\,$kg$^{-1}$K$^{-1}$,  we get  $ \Delta H_T (T_c) = 4.2 \, $kJ/kg or 6.9~meV/FeRh. 
In general, $\Delta H_T \ne \Delta H_P$, but  $\Delta H_P$ is typically measured in experiments at fixed external pressure $P$.

\subsection*{Transition Temperature $T_c$ }
We note that transition temperature $T_c$ in eq.~\ref{eDE} can be estimated accurately in mean-field approximations but only if considered separately for segregating (immiscible) \cite{PRB75p104203y2007} and ordering (miscible) systems \cite{2010.PRB82.024435.CoPt}, which have a negative formation enthalpy, e.g., stable solid-solution phase.

{\par} For a segregating system, a mean-field approximation was shown to be highly accurate for miscibility gaps (the so-called $T_{0}$ line)  
away from compositional limits (i.e.,  $c_{\alpha} \rightarrow 0$ or $c_{\alpha} \rightarrow 1$ for an $\alpha$ atomic type), 
where mean-field entropy differences are less accurate. (Careful Monte Carlo simulations were used to confirm the accuracy \cite{PRB75p104203y2007}.) 
However, in these cases, vibrational entropy changes can have a large effect in  $T_c$, where analytically it is changed when going between two phases (e.g., solid solution and segregation)  as
\begin{equation}
\label{eTc-Vibs}
    T_c = T_{c,conf} \left(1 + \frac{ \Delta S_{L} } { \Delta S_{conf} } \right)^{-1}, 
\end{equation}
where the subscript  ``conf'' delineates the configurational entropy only and $\Delta S_{L}$ is the lattice vibrations entropy changes. 
Moreover, to a good approximation (at least in binary metals), \cite{Bognanoff-Fultz1999,Delaire-Fultz2006}
 the lattice vibrational change $\Delta S_L \approx -0.34 \Delta\chi$, where $\Delta\chi$ is the electronegativity difference between alloying pairs. So, if the electronegativities of elemental pairs are similar, there is no effect from vibrations on $T_c$ and estimates without vibrational calculations are fine, as discussed in Ref.~\onlinecite{PRB75p104203y2007}. Otherwise, changes in vibrational entropy can be estimated at a given temperature, as we have outlined earlier. 

{\par} Typically, the sign of a formation enthalpy $H_f$ indicates either segregation ($H_f>0$) or ordering ($H_f \le 0$) tendency. 
Any diffusion broadens the hysteresis, while a chemical inhomogeneity smears a diffusion\-less phase transition; 
both effects are consequences of a segregation tendency, which should be avoided in caloric materials. 
Fortunately, a positive formation enthalpy can easily be monitored during materials screening.

{\par} In contrast to segregation for miscible alloys (e.g., FeRh systems exemplified here), a $T_c$ estimate for a first-order transition between two phases can be estimated well by
\begin{equation}
\label{eTc}
    T_c = I_c \cdot \delta H_0 / k_B , 
\end{equation}
where $\delta H_0$ is the enthalpy difference between fully-relaxed structures at zero temperature, 
and $I_c \sim 1$ (dimensionless) is a factor with a constant value for a class of similar systems.
Please keep in mind that $\delta H_0$ should not be confused with $\Delta H_T (T_c)$, and, as expected, 
$ \Delta H_T (T_c) \ll \delta H_0 (0 \, \mbox{K})$, as numerically exemplified before \cite{ZarkevichPhD2003}.
We have found that eq.~\ref{eTc} accurately estimates order-disorder transitions in metallic alloys  \cite{ActaMat50n9p2443y2002,PRB67p064104y2003} and  
martensitic phase transitions \cite{PRB90p060102y2014,PRL113p265701y2014,2010.PRB82.024435.CoPt}.

{\par}Equation \ref{eTc} with $I_c = 1$ is exact for barrierless transitions, whereas generally $I_c$ is proportional to a ratio (of functions of order parameters) nearing $1$ between the two systems, such as two magnetic configurations in a fixed chemical cell or in an order-disorder transitions in a fixed magnetic state \cite{2010.PRB82.024435.CoPt}.
For example, the calculated enthalpy difference $\delta H_0$ between   AFM  and FM  B2-FeRh is $29.8 \pm 1\,$meV/atom (Fig.~\ref{figEPV}); this value compares well with previous calculations \cite{PRB46p2864y1992}.
For the meta\-magnetic phase transition in FeRh, we find that $( \delta H_0 / k_B ) = 346 \pm 12 \,$K, which compares well with $T_c = 353 \pm 1 \,$K measured in Fe$_{50}$Rh$_{50}$ \cite{PhysRev131p183y1963}.
The value of $I_c$ near $1$  has uncertainty due to an error in DFT energies and in the measured $T_c$.
As the chemical structure is fixed for FeRh metamagnetic transition and only the magnetic configuration changed,  it is purely an electronic configurational change.

{\par} Equations~\ref{eDE} and \ref{eTc} are exact, while $I_c \approx 1$ is approximate.  
For barrierless transitions, the enthalpy difference $\delta H_0$ coincides with the energy needed to excite an additional degree of freedom (DoF) 
and access the higher-temperature phase, and in the classical limit $I_c \equiv 1$ in this case.  
This interpretation of eq.~\ref{eTc} was successfully applied to estimate melting temperatures  \cite{PRL100p040602}.
 The apparent simplicity of the estimate (\ref{eTc}) obscures a complicated counting of the number of the effective DoF  \cite{PRL100p040602}.
 In general, a higher-$T$ phase has more DoF contributing and consequently a higher entropy than the lower-$T$ phase. 
 The change in the number of effective DoF is an integer, hence, a reasonable accuracy of the eq.~(\ref{eTc}) with $I_c \approx 1$ is not a coincidence.
 As both atomic and spin orderings can be described by a basis-set expansion \cite{PRL92p255702y2004},  a similar equation for different physics is obtained. 
 One can assess eq.~\ref{eTc} for generic alloy screening, as exemplified for order-disorder transitions in Table~\ref{T1TH} or for solid-solid phase transitions in Table \ref{T3Tc}.

\begin{table}[h!]
\centering
\begin{tabular}{ccccccll}
\hline
System & GS & $\delta H_0$  & $k_B T_c$ & $T_c$  & Expt. & $I_c^{-1}$ & Ref. \\
             &       & (meV)    &  (meV)       &  (K)  &  (K)  &  &  \\
\hline
 Ag$_3$Al & ($D0_{22}$) & 46 & 45 &  520 & - & 0.98 & \cite{JohnsonAsta1997,AstaJohnson1997,AstaHoyt2000} \\   
 Ag$_2$Al & (MoPt$_2$)  & 41 & 37 & 430 & - & 0.91 & \cite{JohnsonAsta1997,AstaJohnson1997,AstaHoyt2000} \\
\hline
  AgAu & [$L1_0$]            & 12.2  & 14 & 165 & - & 1.13 & \cite{PRB45n2p613y1992} \\  
  AgAu & [$L1_0$]            & 16.7 & - & - & - & - & \cite{PRB57p6427y1998} \\  
\hline
 CuAu &  $L1_0$             & 47  & 48.3 & 560 & 658 & 1.03 & \cite{PRB58pR5897y1998} \\  
 Cu$_3$Au & [$L1_2$]   & 42.8 & - & - & 500 & - & \cite{PRB57p6427y1998} \\
\hline
 Ni$_3$V & $D0_{22}$  &  115 & 118 & 1370 & 1318 & 1.03 & \cite{PRL92p255702y2004} \\
\hline
\end{tabular}
\caption{\label{T1TH}
Calculated order-disorder enthalpies $\delta H_0$ (meV/atom) and $k_B T_c$ (meV) and $T_c$ (K) for some fcc (miscible) binaries, using our data from Table 3.4 in Ref.~\onlinecite{ZarkevichPhD2003}.
Strukturbericht designations in brackets [...] are assumed ground states structures; structures in parenthesis (...) are metastable. 
Calculated $T_c$'s were obtained from Monte Carlo using a cluster expansion, and compared to experimental (Expt.) values \cite{Hultgren1963,Okamoto2010}.
}
\end{table}

\begin{table}[h!]
\centering
\begin{tabular}{ccccccll}
\hline
System & Phase & Transition & $\delta H_0$  & $\delta H_0 / k_B$ & $T_c$  & $I_c^{-1}$ & Ref. \\
             &       &                 & (meV)             &  (K)       &  (K)  &   &   \\
\hline
Ti & hcp & hcp-bcc &  97.2 & 1128 & 1155 & 0.976 & present \\
Hf & hcp & hcp-bcc & 174.2 & 2022 & 2016 & 1.003 & present \\
NiTi & bco & martensitic & 29.5 & 343 & 333 & 1.030 & \cite{PRL113p265701y2014} \\   
FePd & $L1_0$ & FM-PM & 64.5 & 749 & 730 & 1.026 & \cite{PRB82p024435y2010} \\
FePt &  $L1_0$ & FM-PM &  63.3 & 735 & 750 & 0.980  & \cite{PRB82p024435y2010} \\
CoPt & $L1_0$ & FM-PM &  59.6 & 692 & 720 & 0.961 & \cite{PRB82p024435y2010} \\
MnNiSi & $Pnma$ & FM-PM & 55.7 & 646 & 662 & 0.976 & present \\
LiBH$_4$ & $Pnma$ & melting & 45.3 & 526 & 553 & 0.951 & \cite{PRL100p040602} \\
FeRh & B2 & AFM-FM & 29.8 & 346 & 353 & 0.980 & present \\
\hline
\end{tabular}
\caption{\label{T3Tc}
Calculated examples of relative enthalpies $\delta H_0$ (meV/atom) between low-T and high-T phases  are compared 
to  experimental $T_c$'s \cite{Okamoto2010,ASM2018apd}, with $I_c^{-1} = \delta H_0 / k_B T_c $ (dimensionless).  
}
\end{table}

\subsection*{Compositional Sensitivity of $T_c$}
Notably, $T_c$ scales with $\delta H_0$ in both stoichiometric  (50~at.\%~Rh) and off-stoichiometric alloys with a partial atomic disorder, including with long-range order parameter, see, e.g., Ref.~\onlinecite{2010.PRB82.024435.CoPt}. 
From the electronic density of states (DOS) $n(E)$ in Fig.~\ref{figDOS}, 
also seen in recent calculations \cite{PRB93p024423y2016,PRB94p014109y2016,PRB94p174435y2016}, 
we expect that lowering of the Fermi energy E$_F$ (due to decrease in Rh fraction)  will stabilize the FM phase (from a lower DOS in the pseudogap), but it would have a lesser effect on the AFM phase.   
This change will decrease $\delta H_0$ and will reduce $T_c$.  
Indeed, this qualitative expectation agrees with the experimental phase diagram \cite{PhysRev131p183y1963,Intermetallics15n9p1237y2007,Swartzendruber1984}.   
Compositional hypersensitivity of FeRh was theoretically studied in Ref.~\onlinecite{PRB89p054427y2014}.

\subsection*{Field Dependence of $T_c$}
Dependence of $T_c$ on the external magnetic field $B$,  
as well as dependence of the critical field $B_c$ on $T$, assuming $(d B_c / dT)^{-1} = dT_c / dB$, 
can be determined from discontinuities in magnetization $M$ and entropy $S$ at the first-order meta\-magnetic transition:
\begin{equation}
\label{dTdB}
\frac{dT_c}{dB} = \frac{\Delta M_{T=T_c}}{ \Delta S_{T=T_c} }  .
\end{equation}
 The calculated magnetizations of the fully-relaxed B2-FeRh in AFM and FM states are $0$ and $2.1 \, \mu_B$/atom, respectively (Section~\ref{atomicM}).  
For the upper bound $\Delta M (T_c) \!<\! [M(\mbox{FM}) \!-\! M(\mbox{AFM})]$ for the magnetization change at $T_c$,    
we find  
$\Delta M / \Delta S_{T} <  2.1 \mu_B / 0.103 k_B = 13.7 \,$K/Tesla.
However, a more realistic value \cite{jjimm80n3p186y2016}  of  $(\Delta M)_{T_c}$ -- 60\%  of the upper bound  -- gives $-dT_c /dB = 8.2\,$K/Tesla for stoichiometric FeRh. 
Measurements of $T_c (B)$ in the external magnetic field  (or critical field vs.~$T$) provide a quadratic \cite{JPhysC3n1SpS46y1970} dependence
with the linear  \cite{JETP19n6p1348y1964} slope $- dT_c /dB$ in small fields of 
$8.2\,$ in FeRh \cite{JAP37n3p1257y1966};   
$8.2\,$ in Fe$_{49.5}$Rh$_{50.5}$ \cite{jjimm80n3p186y2016};   
$8.5\,$ in Fe$_{49}$Rh$_{51}$ \cite{ActaMat106p15y2016}; and
from $9.6$ to $9.7 \,$K/Tesla in Fe$_{49}$Rh$_{51}$ \cite{PRB89p214105y2014}.

\subsection*{Accuracy}
{\par}  As shown, a number of standard approximations within DFT calculations work very well for estimating many thermodynamic quantities, in particular for caloric properties, such as  transition temperatures $T_c$, field-dependent changes in $T_c$,  and electronic entropy changes $\Delta S_e$ (the main contribution), 
while  the significant lattice entropy changes $\Delta S_L$ are underestimated for anharmonic atomic vibrations, 
 which are found in many systems with lattice instabilities. However, we established a direct method to evaluate more correctly $\Delta S_L$, which gave a 50\% increase in its magnitude, and provided more accurate estimates of caloric properties, see Table~\ref{t3}.  
 It remains to test these estimators in more complex systems to screen for improved caloric materials via an approach presented recently \cite{JPhysD51n2p024002y2018}.

\section{\label{Discussion}Issues with Indirect Assessments} 
{\par} Before closing, we would be remiss not to remark on quantities that are difficult to assess theoretically due to errors or inability to measure experimentally,
clearly relevant to materials screening, and occasional incorrectly applied.

{\par} Often  the measured ${dT_c}/{dP}$ and $\Delta V$ is used to evaluate $\Delta S$ using the Clausius-Clapeyron equation 
\begin{equation}
\label{ClausiusClapeyron}
\frac{dT_c}{dP} = \frac{\Delta V_T}{ \Delta S_P }   .
\end{equation}
However, there is a well-known problem with applications of this equation to experimental data \cite{Ricodeau1972}. 
Specifically, while it is possible to measure pressure $P$ and the corresponding volume change $(\Delta V_P)$ at a first-order transition, the isothermal volume change $(\Delta V_T)$ induced by varying $P$ is not measured; and, furthermore, there is no reason that $\Delta V_T$ and  $\Delta V_P$ are the same. 
Nonetheless, there have been instances where $\Delta V_P$ was used as equal to  $\Delta V_T$ to use eq.~\ref{ClausiusClapeyron}, which gives an overestimate of $\Delta S$, see, for example,  Ref.~\onlinecite{JETP19n6p1348y1964}.
Such disagreements  of estimates from eq.~\ref{ClausiusClapeyron} and direct measurements are well documented \cite{Ricodeau1972}.  
Pressure dependence of  $T_c$ has been long discussed \cite{JETP19n6p1348y1964,Ricodeau1972,PhysRev170p523y1968}; 
the measurements  \cite{PhysRev170p523y1968} of ${dT_c}/{dP}$  in FeRh range from 
43 \cite{JETP19n6p1348y1964}  to 
64 K/GPa \cite{PRB89p214105y2014}. 

{\par} Regarding the  accuracy of DFT-calculated energy ($E$) versus volume ($V$) curves (Fig.~\ref{figEPV}), 
the  lattice constants $a_0$ (at $P=0$~GPa, $T=0$~K) are 2.996 {\AA} in AFM and 3.012 {\AA} in FM phase for B2 FeRh,  
while the  measurements on Fe$_{50}$Rh$_{50}$ at $353\,$K give 2.987~{\AA} and 2.997~{\AA},   \cite{PSSB20n1pK25y1967} similar to results in Ref.~\onlinecite{JETP19n6p1348y1964}, see our Table~\ref{t3}. 
So, with calculated lattice constants having a relative error of $\pm 0.3\%$, the calculated volume $V \! \sim \! a^3$  has an error of $\pm 1.2\%$,  too large to determine reliably a change of $\Delta V / V\approx 1\%$, as found relevant in experiment \cite{PSSB20n1pK25y1967}. 
So, one cannot use the Clausius-Clapeyron relation to assess ${dT_c}/{dP}$, if looking for outliers for desired caloric properties.

{\par}Magnetic entropy $S_M$ is typically assessed by thermodynamic integration using experimental data: 
\begin{equation}
 \label{eThermoIntegration}
  \Delta S_M (T, \Delta H) = \int_{H_1}^{H_2} dH  \, \frac{\partial M(T,H)}{ \partial T}|_H  .
\end{equation}
Importantly, this equation is valid within a single phase.  
Derivative $\partial M / \partial T$ diverges at the meta\-magnetic first-order phase transition. 
Thermodynamic integration should not be performed across phase boundaries. 

{\par}In addition, the difference between values in two phases should be calculated by subtracting values obtained for the same chemical composition, 
otherwise improper or misleading results can be derived, as in Ref.~\onlinecite{PRL109p255901y2012}, 
where two epitaxial Fe-Rh films of different compositions we used, i.e., Fe-rich with FM ground state and Rh-rich with AFM ground state.

\section{\label{Discussion2}Generic Remarks}
{\par} 
\subsection{Bounds and Dominant Contributions for Entropy Change}  
For any type of screening, it is useful to note the largest contributions that can be expected to control desired behavior.
For caloric behavior, electronic and lattice entropy changes due to electronic- or structural-driven instabilities are most critical and we can approximate the largest possible values.
Namely, for $d$-band ($f$-band) systems, the electronic spin (magnetic) entropy changes $\Delta S_e \le \Delta S_e^{max}$ have 
upper limit of $\Delta S_e^{max}/k_B = \ln(2^{n/2})$ of $3.47$  ($4.85$) per half-filled band with $n$ being $10$~$d$ ($14$~$f$) orbitals; 
this is essentially the maximum permitted magnetic entropy change from atomic magnetization.  
$\Delta S_e$ cannot be larger than the electronic entropy $S_e$ of either phase, as estimated by eq.~\ref{eSommerfeld}. 
A phase transition accompanied by a large change of electrical conductivity (proportional to electronic DOS at $E_F$, i.e., $n(E_F)$) is expected to have a good $\Delta S_e$. 

{\par} If the transition temperature between competing states is above the respective Debye temperatures, 
the vibrational entropy change for a solid-solid transition is approximated by  
\begin{equation}
 \label{UpperBound-S}
\Delta S_L \approx 3 k_B \ln(\Theta^D_{2}/\Theta^D_{1}) ,
\end{equation}
where $\Theta^D_{\alpha}$ is the Debye temperature of the phase $\alpha$. 
For $\Theta^D_{2}/\Theta^D_{1}$ of $1.0$ to $1.5$, a safe upper-bound range for solids of the same stoichiometry and pressure, 
we get  $0 \le \Delta S_L  < 1.22 \, k_B$/atom for quasi\-harmonic solids,  a bound smaller than that for electronic contributions 
(i.e., $\Delta S_L  < \Delta S_e$).    
Also, $\Delta S_L$ in a solid with non-harmonic phonons can be larger than that in a harmonic solid, as already demonstrated.
The general expectation then is that the combined electronic and magnetic entropy changes will 
constitute the dominant contributions to the total  $\Delta S_T$ for caloric systems, 
while the lattice entropy can be significant but secondary (and more demanding to estimate reliably). 
An estimate of the dominant effect (and its bounds) is used for the high-throughput pre-screening of materials \cite{Complexity11p36y2006,JPhysD51n2p024002y2018}.

\subsection{Chemical Disorder and Segregation} 
Caloric material is expected to have a phase transition at the target temperature $T$.  
However, stoichiometric line compounds typically have off-target values of $T_c$.  
To correct this, chemical composition is altered and an off-stoichiometric chemical disorder is introduced. 
For a large caloric effect, the first-order phase transition must be sharp, 
and consequently the caloric material must be chemically homogeneous.  
Any segregation will be detrimental to such homogeneity. 

{\par} To screen out segregating materials, we use the coherent-potential approximation (CPA), \cite{CPA2} implemented in the KKR electronic-structure code, \cite{MECCA}  
to compute dependences of the formation enthalpy $H_f$ on composition $c$, considering possible disorder on each sub\-lattice. 
(One can also use large representative  super\-cells at a number of discrete compositions, 
but, if done carefully, those results usually compare well with the output of KKR-CPA, which is much faster to compute due to smaller cells with fewer atoms and electrons.)
If immiscible, i.e., $\partial^2 H_f / \partial c^2 < 0$, then $H_f (c)$ is concave and the system can lower its energy by developing a compositional inhomogeneity (segregation) that is unfavorable for calorics. 
Such materials are rejected, such as (Hf$_{1-c}$Nb$_c$)Fe$_2$ Lave's phase (Fig.~9 in Ref.~\onlinecite{ActaMat154p365y2018}). 
In contrast, a convex $H_f (c)$  (in miscible system with $\partial^2 H_f / \partial c^2  > 0$) is a necessary but not sufficient condition for good caloric properties.
An example with a convex $H_f (c)$ is ZrMn$_6$(Sn$_{1-c}$Sb$_c$), see Fig.~8 in Ref.~\onlinecite{JPhysD51n2p024002y2018}. 

\subsection{Hysteresis}
A first-order phase transition is usually accompanied by a hysteresis. 
The width of the hysteresis serves as huge loss factor for caloric cooling, unless the hysteresis can be eliminated \cite{EnergyTechnology6n8p1397y2018}. 
Nucleation, lattice mismatch, and enthalpy barriers for nucleation and phase boundary propagation contribute to the width of the hysteresis. 
Fortunately, we know how to reduce the hysteresis width.  

{\par} Compositional changes affects the lattice constants in each phase. 
The lattice mismatch between austenite and martensite can be made to go to zero and the hysteresis thereby narrowed by the fine tuning of composition $c$ \cite{NatureMater5p286y2006},  
which occurs when the middle eigenvalue ($\lambda_2$) of the transformation stretch tensor attains the value $1$ at $T_c$.
 While $\lambda_2 (c)$ could be monitored versus composition, it is far more convenient and straightforward to assess the dependence of the lattice constants in the relevant phases on composition at fixed $(P, T)$, as computed in DFT, see section~\ref{ComputationalDetails}.  
The KKR-CPA permits to do this easily and quickly for materials with disorder, as we have done many times.  
Typically, only a few calculations are needed to find compositions where lattice match is achieved.

{\par}  Finally, defects (e.g., surface geometry,  bulk impurities, precipitates, or  second-phase remnants due to incomplete transformation) can serve as nucleation centers, suppressing the nucleation enthalpy barriers.   Design of caloric devices should account for the nucleation centers in caloric materials. 
The enthalpy barriers for the phase boundary propagation depend on composition.  
We calculate them using the nudged elastic band (NEB) methods \cite{NEB1998,SSNEB,C2NEB}. 
Unfortunately, $T_c$ depends on composition, too.  
Hence, reduction of the hysteresis at constant $T_c$ by adjusting $c$ is similar to tuning a piano: 
several compositional degrees of freedom must be simultaneously or iteratively adjusted to get the target values for both $T_c$ and hysteresis width. 
Nevertheless, trends can be assessed with relatively few calculations to find better design regions, or eliminate systems quickly \cite{JPhysD51n2p024002y2018}.

\section{\label{Beyond}Beyond FeRh:  Novel Materials with Giant Magnetcaloric Response}
Recently,  we utilized some of these methods to search among $10,000+$ candidates and to reduce systems of interest for our experimental collaborators, eliminating thousands alloys \cite{JPhysD51n2p024002y2018}. 
Out of all systems scanned, about ten  (or 0.1\%) were found ("predicted") to have caloric behavior either similar to FeRh (i.e., $\Delta S_T \ge 12\,\,\mbox{J}\,\mbox{kg}^{-1}\mbox{K}^{-1}$), but lower in cost or significantly improvable with modifications to alloying chemistry. 
Several classes of these materials are now being investigated experimentally.  
For example, Ni-Co-Mn-Ti \cite{Bez2019} and Mn$_{0.5}$Fe$_{0.5}$NiSi$_{1-x}$Al$_x$ \cite{Biswas2019} have been confirmed to be  promising for the solid-state refrigeration, with  enhancement of $\Delta S_T$ well above $20\,\,\mbox{J}\,\mbox{kg}^{-1}\mbox{K}^{-1}$ at room temperature.
Such discovery will be accelerated when this type of screening is implemented through a database combined with key correlations derived by machine-learning techniques, especially when looking for outliers in desired properties -- just as with systems with zero hysteresis at phase transformations, where the desired compositional range may consist of a single point \cite{NatureMater5p286y2006}.

\section{\label{Summary}Summary}
{\par} We have explored several thermodynamic estimates for assessing caloric properties in alloys.
We used FeRh as a testbed, as it exhibits large multicaloric (magneto-, elasto- and baro-caloric) responses at its meta\-magnetic transition just above room temperature, as well as non-harmonic  vibrations -- typical for systems near lattice instabilities. 
We showed that use of controlled $T$-dependent atomic displacements, easily estimated at $T_c$,  provides a reliable assessment of lattice entropy changes at the phase transition. 
In FeRh, we tested approximate methods and estimators, and evaluated a number of thermodynamic properties, including specific heat, entropy and enthalpy changes, transition temperature, and isentropic temperature drop.  
The predicted caloric properties are in a quantitative agreement with the trusted experimental data, see Table~\ref{t3}. We have verified that these estimators are reliable (if applied carefully) and accurate. 
In contrast, we showed that some previously used assessments, like from the Clausius-Clapeyron relation (\ref{ClausiusClapeyron}), are unreliable due to the underlying assumptions.
Thus, assessment and testing of the methods were a necessity. 

{\par } Tested reliable methods will enable faster theory-guided screening to find more promising caloric materials, involving more complex multicomponent systems on which to focus.  Indeed, the estimates provided here already resulted in finding improved lower-cost caloric systems exhibiting giant magnetocaloric enhancements with promise for use in solid-state cooling \cite{Bez2019,Biswas2019}.


\section*{Acknowledgments}
We thank Dr. Vitalij Pecharsky and Dr. Klaus Ruedenberg for discussions. 
Our theory developments at Ames Laboratory and Iowa State University were funded by the U.S. Department of Energy (DOE),  Office of Science, Basic Energy Sciences, Materials Science and Engineering Division.  Ames Laboratory is operated for the U.S. DOE by Iowa State University under contract DE-AC02-07CH11358.
Initial application to caloric materials discovery coupled with experimental studies was partly supported by the U.S. DOE,  Advanced Manufacturing Office of the Office of Energy Efficiency and Renewable Energy through CaloriCool\textsuperscript{TM} -- the Caloric Materials Consortium established as a part of the U.S. DOE Energy Materials Network \cite{EnergyMaterialsNetwork}.

\bibliography{FeRh}
\end{document}